\documentclass[letterpaper,10pt,conference]{IEEEtran}
\IEEEoverridecommandlockouts % enable \thanks

\makeatletter
\newcommand{\AddInputPath}[1]{%
  \ifx\input@path\@undefined
    \def\input@path{#1}
  \else
    \g@addto@macro{\input@path}{#1}
  \fi
}
\makeatother

\usepackage{shellesc}

\usepackage{etex}

\usepackage[font=footnotesize,caption=false]{subfig} % preload with correct options...

\usepackage{relsize}

\usepackage[dvipsnames,svgnames,usenames]{xcolor}
\usepackage{array}
\usepackage{booktabs,tabularx}
\usepackage{multirow}
\usepackage[binary-units=true,per-mode=symbol,detect-mode=true]{siunitx}
\usepackage{graphicx}

\usepackage[T1]{fontenc}
\usepackage{textcomp}
\usepackage[utf8]{inputenc}
\usepackage{icomma}
\usepackage{xspace}

\usepackage[tbtags]{amsmath}
\usepackage{amssymb,amsfonts,bm}
\usepackage{mathtools} 
\usepackage{dsfont}
\usepackage{mathrsfs}
\usepackage{accents}
\usepackage{empheq}
\usepackage{nccmath}
\usepackage{balance}
\usepackage{setspace}

\usepackage{color}
\usepackage{calc}
\usepackage{tikz}
\usepackage{pgfplots,pgfplotstable}

\usepackage{pdftexcmds}
\makeatletter
\newcommand{\strequal}[2]{\pdf@strcmp{#1}{#2}==0}
\makeatother

\usepackage[final,bookmarksnumbered=true,pdfborder={0 0 0}]{hyperref}
\usepackage[capitalize]{cleveref}
\usepackage{refcount}

\usepackage[inline]{enumitem}
\usepackage{algorithm}
\usepackage{algpseudocode}
\makeatletter
\newcommand{\algmargin}{\the\ALG@thistlm}
\algblockx{StartIA}{EndIA}[1]{\textbf{Input} #1}[1]{\textbf{Return} #1}
\makeatother
\newlength{\whilewidth}
\algdef{SE}[parWHILE]{parWhile}{EndparWhile}[1]
  {\parbox[t]{\dimexpr\linewidth-\algmargin}{%
     \hangindent\whilewidth\strut\algorithmicwhile\ #1\ \algorithmicdo\strut}}{\algorithmicend\ \algorithmicwhile}%
\algnewcommand{\parState}[1]{\State%
  \parbox[t]{\dimexpr\linewidth-\algmargin}{\strut #1\strut}}

\makeatletter
\newcommand\fs@spaceruled{\def\@fs@cfont{\bfseries}\let\@fs@capt\floatc@ruled
  \def\@fs@pre{\vspace{.05in}\hrule height.8pt depth0pt \kern2pt}%
  \def\@fs@post{\kern2pt\hrule\relax}%
  \def\@fs@mid{\kern2pt\hrule\kern2pt}%
  \let\@fs@iftopcapt\iftrue}
\makeatother

\floatstyle{spaceruled}% Select new float style
\restylefloat{algorithm}% Apply spaceruled float style to algorithm

\usepackage{glossaries-prefix}
\usepackage{ifthen}
\usepackage[noadjust]{cite}
\usepackage{multibib}

\usepackage{comment}
\usepackage{todonotes}
\let\legacytodo\todo
\newcommand{\ruggedtodo}[2][]{\tikzexternaldisable\legacytodo[#1]{#2}\tikzexternalenable}
\renewcommand{\todo}[1]{\ruggedtodo[inline]{#1}}

\makeglossaries

\newacronym{lln}{LLN}{law of large numbers}
\newacronym{soa}{SoA}{state-of-the-art}
\newacronym{mh}{MH}{multi-hop}
\newacronym{tcs}{TCS}{time correlated sparsification}
\newacronym{ia}{IA}{incremental aggregation}
\newacronym{sfl}{SFL}{satellite federated learning}
\newacronym{esa}{ESA}{European Space Agency}
\newacronym{tle}{TLE}{two-line element set}
\newacronym{ai}{AI}{artificial intelligence}
\newacronym{ann}{ANN}{artificial neural network}
\newacronym{jscc}{JSCC}{joint source-channel coding}
\newacronym{raan}{RAAN}{right ascension of the ascending node}
\newacronym{uav}{UAV}{unmanned aerial vehicle}
\newacronym{haps}{HAPS}{high-altitude platform station}
\newacronym{6g}{6G}{sixth generation}
\newacronym{cgr}{CGR}{contact graph routing}
\newacronym{dtn}{DTN}{delay-tolerant networking}
\newacronym{fl}{FL}{federated learning}
\newacronym{dl}{DL}{deep learning}
\newacronym{fedavg}{FedAvg}{federated averaging}
\newacronym{dml}{DML}{distributed ML}
\newacronym{rl}{RL}{reinforcement learning}
\newacronym{ps}{PS}{parameter server}
\newacronym[prefix={an\space},prefixfirst={a~}]{ml}{ML}{machine learning}
\newacronym{sgd}{SGD}{stochastic gradient descent}
\newacronym{dsgd}{DSGD}{distributed stochastic gradient descent}
\newacronym{isl}{ISL}{inter-satellite link}
\newacronym{gsl}{GSL}{ground to satellite link}
\newacronym{gs}{GS}{ground station}
\newacronym{ecef}{ECEF}{earth-centered, earth-fixed}
\newacronym{eci}{ECI}{Earth-centered inertial}
\newacronym{ofdm}{OFDM}{orthogonal frequency-division multiplexing}
\newacronym{cp}{CP}{cyclic prefix}
\newacronym{los}{LOS}{line-of-sight}
\newacronym{leo}{LEO}{low earth orbit}
\newacronym{meo}{MEO}{medium earth orbit}
\newacronym{gso}{GSO}{geosynchronous orbit}
\newacronym{geo}{GEO}{geostationary}
\newacronym{eo}{EO}{Earth observation}
\newacronym{iot}{IoT}{Internet of Things}
\newacronym{irs}{IRS}{intelligent reflecting surface}
\newacronym{socp}{SOCP}{second-order cone program}
\newacronym{soc}{SOC}{second-order cone}
\newacronym{dsl}{DSL}{digital subscriber line}
\newacronym{wsee}{WSEE}{weighted sum energy efficiency}
\newacronym{mmwave}{mmWave}{millimeter wave}
\newacronym{dfg}{DFG}{Deutsche Forschungsgemeinschaft}
\newacronym{haec}{HAEC}{Highly Adaptive Energy-Efficient Computing}
\newacronym{hpc}{HPC}{High Performance Computing}
\newacronym{mac}{MAC}{multiple-access channel}
\newacronym{bc}{BC}{broadcast channel}
\newacronym{siso}{SISO}{single-input single-output}
\newacronym{simo}{SIMO}{single-input multiple-output}
\newacronym{miso}{MISO}{multiple-input single-output}
\newacronym{mimo}{MIMO}{multiple-input multiple-output}
\newacronym{af}{AF}{amplify-and-forward}
\newacronym{df}{DF}{decode-and-forward}
\newacronym{cf}{CF}{compress-and-forward}
\newacronym{mwrc}{MWRC}{multi-way relay channel}
\newacronym{dmmwrc}{DM-MWRC}{discrete memoryless multi-way relay channel}
\newacronym{pde}{PDE}{partial data exchange}
\newacronym{fde}{FDE}{full data exchange}
\newacronym{iid}{i.i.d.\@}{independent and identically distributed}
\newacronym{di}{DI} {difference of increasing}
\newacronym{dc}{DC}{difference of convex}
\newacronym{mm}{MM}{mixed monotonic}
\newacronym{mmp}{MMP}{mixed monotonic programming}
\newacronym{awgn}{AWGN}{additive white Gaussian noise}
\newacronym{wgn}{WGN}{white Gaussian noise}
\newacronym{awg}{AWG}{additive white Gaussian}
\newacronym{sic}{SIC}{successive interference cancellation}
\newacronym{snr}{SNR}{signal-to-noise ratio}
\newacronym{sinr}{SINR}{signal to interference plus noise ratio}
\newacronym{inr}{INR}{interference to noise ratio}
\newacronym{zf}{ZF}{zero-forcing}
\newacronym{mrt}{MRT}{maximum ratio transmission}
\newacronym{mmse}{MMSE}{minimum mean square error}
\newacronym{sud}{SUD}{single user decoding}
\newacronym{dof}{DoF}{degrees of freedom}
\newacronym{gdof}{GDoF}{generalized degrees of freedom}
\newacronym{nnc}{NNC}{noisy network coding}
\newacronym{dmn}{DMN}{discrete memoryless network}
\newacronym{csi}{CSI}{channel state information}
\newacronym{pmf}{pmf}{probability mass function}
\newacronym{dmic}{DM-IC}{discrete memoryless interference channel}
\newacronym{ic}{IC}{interference channel}
\newacronym{gic}{GIC}{Gaussian interference channel}
\newacronym{if}{IF}{interference}
\newacronym{ee}{EE}{energy efficiency}
\newacronym{gee}{GEE}{global energy efficiency}
\newacronym{tin}{TIN}{treating interference as noise}
\newacronym{snd}{SND}{simultaneous non-unique decoding}
\newacronym{sd}{SD}{simultaneous decoding}
\newacronym{hk}{HK}{Han-Kobayashi}
\newacronym{rs}{RS}{rate splitting}
\newacronym{rf}{RF}{radio frequency}
\newacronym{pa}{PA}{power amplifier}
\newacronym{lna}{LNA}{low noise amplifier}
\newacronym{lo}{LO}{local oscillator}
\newacronym{adc}{ADC}{analog-to-digital converter}
\newacronym{dac}{DAC}{digital-to-analog converter}
\newacronym{dsp}{DSP}{digital signal processing}
\newacronym{brd}{BRD}{best response dynamics}
\newacronym{br}{BR}{best response}
\newacronym{ne}{NE}{Nash equilibrium}
\newacronym{lhs}{LHS}{left-hand side}
\newacronym{rhs}{RHS}{right-hand side}
\newacronym{ran}{RAN}{radio access network}
\newacronym{qos}{QoS}{Quality of Service}
\newacronym{ngmn}{NGMN}{Next Generation Mobile Networks}
\newacronym{cap}{CAP}{Capacity Adaptation}
\newacronym{bwa}{BW}{Bandwidth Adaptation}
\newacronym{prb}{PRB}{physical resource block}
\newacronym{se}{SE}{spectral efficiency}
\newacronym{tp}{TP}{throughput}
\newacronym{bs}{BS}{base station}
\newacronym{ue}{UE}{user equipment}
\newacronym{mop}{MOP}{multi-objective optimization problem}
\newacronym{gda}{GDA}{generalized Dinkelbach's algorithm}
\newacronym{midcp}{MIDCP}{mixed integer disciplined convex programming}
\newacronym{lp}{LP}{linear program}
\newacronym{brb}{BRB}{branch reduce and bound}
\newacronym{bb}{BB}{branch and bound}
\newacronym{sit}{SIT}{successive incumbent transcending}
\newacronym{oma}{OMA}{orthogonal multiple access}
\newacronym{noma}{NOMA}{non-orthogonal multiple access}
\newacronym{wlog}{w.l.o.g.\@}{without loss of generality}
\newacronym{lsc}{l.s.c.\@}{lower semi-continuous}
\newacronym{usc}{u.s.c.\@}{upper semi-continuous}
\newacronym{kkt}{KKT}{Karush-Kuhn-Tucker}
\newacronym{ptp}{PTP}{point-to-point}
\newacronym[longplural={Markov decision processes}]{mdp}{MDP}{Markov decision process}
\newacronym[longplural={partially observable Markov decision processes}]{pomdp}{POMDP}{partially observable Markov decision process}
\newacronym{lm}{LM}{learning machine}
\newacronym{llm}{LLM}{large language model}
\newacronym{erm}{ERM}{empirical risk minimization}
\newacronym{dqn}{DQN}{Deep Q-Network}
\newacronym{ppo}{PPO}{Proximal Policy Optimization}
\newacronym{tl}{TL}{transfer learning}
\newacronym{gpt}{GPT}{generative pre-trained transformer}
\newacronym{agi}{AGI}{artificial general intelligence}
\newacronym{ood}{OOD}{out of distribution}
\newacronym{nist}{NIST}{National Institute of Standards and Technology}
\newacronym{itu-r}{ITU-R}{International Telecommunication Union, Radiocommunication Sector}
\newacronym{us}{U.S.\@}{United States}
\newacronym{urllc}{URLLC}{ultra reliable low latency communications}
\newacronym{urc}{URC}{ultra reliable communications}

\glsenableentrycount
\makeglossaries

\usetikzlibrary{positioning}
\usetikzlibrary{calc}
\usetikzlibrary{math}
\usetikzlibrary{fit}
\usetikzlibrary{intersections}
\usetikzlibrary{decorations.pathreplacing}
\usetikzlibrary{decorations.markings}
\usetikzlibrary{3d,angles}
\usetikzlibrary{arrows.meta}
\usetikzlibrary{shapes.geometric}
\usetikzlibrary{spath3}

\pgfdeclarelayer{background}
\pgfsetlayers{background,main}

\usetikzlibrary{external}
\usepgfplotslibrary{colorbrewer}

\crefname{equation}{}{}
\crefrangeformat{equation}{(#3#1#4)--(#5#2#6)}
\crefmultiformat{equation}{(#2#1#3)}{ and~(#2#1#3)}{, (#2#1#3)}{, (#2#1#3)}
\crefrangemultiformat{equation}{#3(#1)#4--#5(#2)#6}{, #3(#1)#4--#5(#2)#6}{, #3(#1)#4--#5(#2)#6}{, #3(#1)#4--#5(#2)#6}
\crefrangeformat{algorithm}{Algorithms~#3#1#4--#5#2#6}

\undef\mod
\DeclareMathOperator\mod{mod}

\allowdisplaybreaks[3]

\DeclareSIUnit \dBm {dBm}
\DeclareSIUnit \dBW {dBW}
\DeclareSIUnit \bpcu {bpcu}

\DeclareFontFamily{U}{mathx}{\hyphenchar\font45}
\DeclareFontShape{U}{mathx}{m}{n}{
      <5> <6> <7> <8> <9> <10>
      <10.95> <12> <14.4> <17.28> <20.74> <24.88>
      mathx10
      }{}
\DeclareSymbolFont{mathx}{U}{mathx}{m}{n}
\DeclareMathSymbol{\bigtimes}{1}{mathx}{"91}

\newtheorem{proposition}{Proposition}

\let\Pr\relax
\DeclareMathOperator{\Pr}{\mathds{P}}

\pgfplotscreateplotcyclelist{default}{%
	blue,mark=*\\%
	red,mark=star\\%
	teal,mark=square*\\%
	brown!60!black,mark=otimes*\\%
}

\newcolumntype{P}[1]{>{\centering\arraybackslash}p{#1}}

\ifCLASSOPTIONdraftcls
\AtBeginEnvironment{figure}{}
\fi

\AtBeginEnvironment{algorithmic}{\footnotesize}

\let\oldFootnote\footnote
\newcommand\nextToken\relax

\renewcommand\footnote[1]{%
    \oldFootnote{#1}\futurelet\nextToken\isFootnote}

\newcommand\isFootnote{%
    \ifx\footnote\nextToken\textsuperscript{,}\fi}

\newcommand\copyrighttext{%
\fontsize{6}{7}\selectfont \textcopyright 2025 IEEE. Personal use of this material is permitted. Permission from IEEE must be obtained for all other uses, including reprinting/republishing this material for advertising or promotional purposes, collecting new collected works for resale or redistribution to servers or lists, or reuse of any copyrighted component of this work in other works.}
\newcommand\copyrightnotice{%
\begin{tikzpicture}[remember picture,overlay]
\node[anchor=south,yshift=20pt] at (current page.south) {\parbox{\textwidth}{\copyrighttext}};
\end{tikzpicture}%
}

\begin{document}
\bstctlcite{IEEEexample:BSTcontrol}
\title{Resilient Radio Access Networks:\\ AI and the Unknown Unknowns}

\author{\IEEEauthorblockN{Bho Matthiesen\IEEEauthorrefmark{1}, Armin Dekorsy\IEEEauthorrefmark{1}, Petar Popovski\IEEEauthorrefmark{2}\IEEEauthorrefmark{1}}
	\IEEEauthorblockA{\IEEEauthorrefmark{1}University of Bremen, Department of Communications Engineering, Germany\\\IEEEauthorrefmark{2}Aalborg University, Department of Electronic Systems, Denmark\\ email: \{matthiesen, dekorsy\}@ant.uni-bremen.de, petarp@es.aau.dk}
\thanks{
This work is supported by the German Research Foundation (DFG) under Grant EXC 2077 (University Allowance).
}%
}
\maketitle
\copyrightnotice

\begin{abstract}
	5G networks offer exceptional reliability and availability, ensuring consistent performance and user satisfaction.
	Yet they might still fail when confronted with the unexpected.
	A resilient system is able to adapt to real-world complexity, including operating conditions completely unanticipated during system design.
	This makes resilience a vital attribute for communication systems that must sustain service in
	scenarios where models are absent or too intricate to provide statistical guarantees.
	Such considerations indicate that artifical intelligence (AI) will play a major role in delivering resilience.
	In this paper, we examine the challenges of designing AIs for resilient radio access networks, especially with respect
	to unanticipated and rare disruptions.
	Our theoretical results indicate strong limitations of current statistical learning methods for resilience
	and suggest connections to online learning and causal inference.
\end{abstract}
\glsresetall

\section{Introduction}
Mobile communication networks are critical infrastructure. Their development has focused on performance, sustainability, reliability, robustness, and security. These capabilities are essential for reliable and efficient operation under a broad range of expected conditions. However, critical infrastructure must also function correctly during extreme events, such as natural disasters.
These events are \emph{rare} and contribute little to a system's overall probability of failure. As a result, they are considered irrelevant in conventional reliability-oriented system design. They are also \emph{unanticipated}, either in the sense that their impact on the system is unforeseeable or their existence is unknown a~priori. This prevents the application of common \cgls{urc} techniques to account for tiny probabilities.
Resilience engineering approaches this challenge through an inherently different perspective on system design.
It focusses on a system's ability to adapt to real world complexity rather than reducing the error probability in expected operating conditions \cite{Madni2009}.
This shifts attention towards a system's capability to cope autonomously with unanticipated conditions, especially those caused by high-impact, low-probability events.

Mobile communication networks were not always seen as critical infrastructure, at least not in the realm of mass-marked commercial networks. This perception changed with increasing technical maturity and widespread deployment over the past decade, especially during the transition from fourth (4G) to fifth (5G) generation systems.
Consequently, resilience has only recently gained recognition as a core requirement for wireless communication networks \cite{NGMN_6G_Resilience,TheWhiteHouse2024,Bennis2024,IMT-2030}.
It is a concept relevant to a great number of scientific disciplines and, while its basic definition is consistent across fields, concrete implications and requirements of resilience vary by system.
A seminal early work on resilience in communication networks is \cite{Sterbenz2010}. Recent works \cite{Khaloopour2024,Mahmood2024,Reifert2024} propose resilience strategies for 6G and adapt the resilience framework in \cite{Sterbenz2010} to contemporary wireless systems. Central to all
is the emphasis on recent advances in \cgls{ai} and \cgls{ml}, sharing an optimistic outlook on their potential for building resilient systems in the near future.

It is undisputed that learning is ``at the heart of resilience'' \cite{Madni2009}. Coupled with the rapid progress in \cgls{ai}, resilience appears, indeed, imminent. However, these \cgls{ai} systems rely almost exclusively on statistical \cgls{ml} methods such as \cgls{dl}. Designed to generalize from samples to distributions, those techniques ``tend to learn statistical regularities in the data set rather than higher-level abstract concepts.''\footnote{This quote is attributed to Turing award winner Yoshua Bengio in \cite{Larson2021}.} This implies an inherent neglect for rare events, a well established property of statistical \cgls{ml} methods \cite{Clemencon2025}.
While not a shortcoming per~se, it appears critical in the context of resilience engineering, which is specifically concerned with rare and unanticipated events.

In this paper, we examine the impact these assumptions have on the design of resilient \cglspl{ran}.
Due to
the lack of established mathematical models for resilience \cite{Bennis2024} and the limited theoretical understanding of \cgls{dl} \cite{Grohs2022}, this is predominantly uncharted territory. In the following, we provide a focused review of resilience in \cref{sec:def} and develop a system model in \cref{sec:sysmod}. Our main results are presented in \cref{sec:sml}, where we highlight several impediments towards the realization of resilient systems with statistical \cgls{ml}. We hope the methodology developed in these sections contributes towards building a deeper mathematical understanding of resilience. This motivates the critical discussion in \cref{sec:discussion}, followed by conclusions in \cref{sec:conclusions}.

\section{Resilience and Unanticipated Challenges} \label{sec:def}
Resilience is a ``semantically overloaded term'' \cite{Madni2009} with varying meaning across different fields.
In general, it is
the ``ability to prepare and plan for, absorb, recover from, or [...] adapt to actual or potential adverse events'' \cite{NationalResearchCouncil2012}.
We focus on wireless communication systems, where the definitions by the \cgls{us}['] \cgls{nist} and the \cgls{itu-r} are of particular relevance.
The \cgls{nist} defines information system resilience as the ability ``to continue to: (i) operate under adverse conditions or stress, even if in a degraded or debilitated state, while maintaining essential operational capabilities; and (ii) recover to an effective operational posture in a time frame consistent with mission needs'' \cite{NIST-SP800-39}. This is well aligned with the \cgls{itu-r} perspective, which considers resilience as the capability of networks and systems ``to continue operating correctly during and after a natural or man-made disturbance, such as the loss of primary source of power, etc.'' \cite{IMT-2030}. A comprehensive survey and discussion of resilience definitions is provided in \cite{Khaloopour2024}.

A valuable supplemental viewpoint is introduced in \cite{Sterbenz2010}. While the author's resilience definition is consistent with those discussed above, they also propose an axiomatic characterization of resilient communication networks. Specifically, Axioms A2 and A3 state that being prepared and having the ability to respond to adverse events are key requirements for resilient systems. The authors emphasize that this includes anticipated and \emph{unanticipated} events of all severities. While anticipated events are predictable based on either past events or a reasoned thread analysis, unanticipated events cannot be predicted ``with any specificity'' \cite{Sterbenz2010}.
In other words, unanticipated events are completely unpredictable during system design and, hence, impossible to address using conventional robust system design methods \cite{Jones2021}.
The ability to cope with such unanticipated failure causes is considered the defining property of resilient systems (over merely robust systems) in \cite{Jones2021}. This perspective is supported by \cite{Jen2005a}, which points out that the inherently stochastic perspective at the core of robust system design does not extend to ``insults previously unencountered and in a real sense unforeseeable.''

Former \cgls{us} Secretary of Defense Donald Rumsfeld popularized the notions of \emph{known unknowns}, that is, things  ``we know [...] we do not know,'' and \emph{unknown unknowns}, which are ``the ones we don't know we don't know'' \cite{UnitedStatesDepartmentOfDefense2002}. This captures the fundamental difference between anticipated and unanticipated events. Thus, the ability to cope with unknown unknowns, created through disruptions in environments that are virtually impossible to predict, is a core capability of resilient systems. Potential sources of such events are mistakes caused by humans, including oversights during system design, natural disasters, political instability, deliberate attacks, and phenomena unknown to science.
Some of these events, sometimes known as 30-sigma events and Black Swans, are extremely rare and occur with probabilities essentially non-computable with scientific methods \cite{Taleb2007}.\footnote{\cGls{urc} is also concerned with extremely rare events. The difference is that \cgls{urc} events are anticipated, while Black Swans are unanticipated.}\footnote{A central thesis of \cite{Taleb2007} is that the probability of rare events cannot be computed by scientific methods due to a lack of data. In essence, the rarer an event is, the less tractable it is and, hence, the less we know about how frequently it might occur \cite{Taleb2016}.}
An immediate consequence is that, during system design, ``all recognized assumptions should be criticized and alternative scenarios developed under the assumption that the assumptions are violated'' \cite{Jones2021}. This implies designing systems under a minimal number of maximally generic assumptions on the operating environment and explicitly ensuring functionality when those assumptions are violated. An interesting observation is that
this design philosophy is antipodal to data-driven systems trained from historically observed or synthetically generated data.
Another implication is that resilience strategies primarily relying on creating a list of potential damaging events from explicit threat analyses are unlikely to succeed.

\section{System Model \& Problem Statement} \label{sec:sysmod}
A \cgls{ran} is the part of a mobile communication system responsible for connecting \cglspl{ue} wirelessly to a core network. This can be modeled as multiple decision-making agents interacting through a shared environment. Here, we are interested in a single agent (or system) $S$ and consider all other agents to be part of system $S$'s environment $E$.  For wireless communications, this environment includes the physical world and, hence, is partially observable, dynamic, nondeterministic, continuous, and partially unknown.

The system $S$ operates in discrete time $t \in \mathds N$ with uniform sampling interval $\Delta_\tau$. That is, we consider time instances $\{ \tau_0 + \Delta_\tau t \}_{t\in\mathds N}$, where $t = 0$ corresponds to the  real time $\tau_0 \in \mathds R$.
For each $t$, the system performs an action $A_t \in \mathcal A$, potentially altering the environment. For a sufficiently large state space $\mathcal E$, the environment $E$ is a Markov decision process, where the state $E_{t+1}$ at time $\tau_0 + \Delta_\tau t$ is determined by conditional transition probabilities $\Pr(E_{t+1} | E_{t}, A_t)$. Accounting for partial observability and physical limitation of sensors, the system only receives an observation $O_t \in \mathcal O$ of $E_t$, where the observation space $\mathcal O$ is assumed significantly smaller than $\mathcal E$ \cite{Javed2024}. For example, a reasonable assumption would be that $\mathcal O$ is countable finite while $\mathcal E$ is uncountable. Based on this observation, and potentially the complete past observation-action-reward history, a supervisor (or teacher) computes a reward $R_t \in \mathcal R$. The observation-reward tuple $(O_t, R_t)$ is input to the system $S$ and serves to determine the action $A_t$. The objective of the system is to choose $A_t$ such that it maximizes the expected reward. To this end, it implements a sequence of measurable mappings $F_t$, $t\in\mathds N$, from $\mathcal A^{t-1} \times \mathcal O^t \times \mathcal R^t$ to $\mathcal A$, potentially taking into to account the complete state-action trajectory $H_t = ( A_0, O_0, R_0, A_1, O_1, R_1, \dots, O_t, R_t )$ up to $t$.

\begin{figure}
	\centering
	\begin{tikzpicture}
		\begin{axis}[
				thick,
				axis lines = center,
				xlabel style = {anchor=north east},
				ylabel near ticks,
				xlabel={Time},
				ylabel={System functionality [$\%$]},
				no markers,
				ymax = 130,
				ymin = 0,
				xmin = 0,
				xmax = 13,
				xtick = {2},
				xticklabels = {},
				ytick = {100,50},
				width=\axisdefaultwidth,
				height=.8*\axisdefaultheight,
				font = \footnotesize,
				clip = false,
				axis on top = true,
				name = plot,
				legend image post style = {thick},
				legend entries = {Conventional response, Resilient response},
				legend columns = 2,
				legend style = {at = {(.5, 1.05)}, anchor = south, cells = {align=left, anchor=west}, /tikz/column 2/.style = {column sep = 1em}},
				set layers,
			]

			\draw[name path=event, line width = 1.5pt] (axis cs:2.001,0) -- (axis cs:2.001,125);

			\node[anchor = north west, inner sep = 0, fill = white] at ($(axis cs:2.001,125)+(.3333em,0)$) {Resilience event $D$};

		% nominal performance
			\addplot[forget plot, black!40, thin, densely dashed] coordinates {
				(1, 100)
				(12.5,100)
			};

		% robust
      \addplot [] coordinates { % resilienceplot.py
            (0, 100) (1.9, 100) (1.90702, 99.9811) (1.91405, 99.9253) (1.92107, 99.8341)
            (1.92809, 99.709) (1.93512, 99.5512) (1.94214, 99.3624) (1.94916, 99.1439)
            (1.95619, 98.8971) (1.96321, 98.6235) (1.97023, 98.3246) (1.97726, 98.0017)
            (1.98428, 97.6564) (1.9913, 97.2899) (1.99833, 96.9039) (2.00535, 96.4996)
            (2.01237, 96.0786) (2.0194, 95.6423) (2.02642, 95.1921) (2.03344, 94.7294)
            (2.04047, 94.2557) (2.04749, 93.7725) (2.05452, 93.2811) (2.06154, 92.783)
            (2.06856, 92.2797) (2.07559, 91.7725) (2.08261, 91.2629) (2.08963, 90.7524)
            (2.09666, 90.2423) (2.10368, 89.7341) (2.1107, 89.2285) (2.11773, 88.7255)
            (2.12475, 88.225) (2.13177, 87.7272) (2.1388, 87.2319) (2.14582, 86.7391)
            (2.15284, 86.249) (2.15987, 85.7613) (2.16689, 85.2763) (2.17391, 84.7938)
            (2.18094, 84.3138) (2.18796, 83.8364) (2.19498, 83.3614) (2.20201, 82.8891)
            (2.20903, 82.4192) (2.21605, 81.9518) (2.22308, 81.487) (2.2301, 81.0246)
            (2.23712, 80.5648) (2.24415, 80.1075) (2.25117, 79.6526) (2.25819, 79.2002)
            (2.26522, 78.7503) (2.27224, 78.3029) (2.27926, 77.8579) (2.28629, 77.4154)
            (2.29331, 76.9754) (2.30033, 76.5378) (2.30736, 76.1026) (2.31438, 75.6699)
            (2.3214, 75.2396) (2.32843, 74.8118) (2.33545, 74.3863) (2.34247, 73.9633)
            (2.3495, 73.5427) (2.35652, 73.1245) (2.36355, 72.7087) (2.37057, 72.2954)
            (2.37759, 71.8844) (2.38462, 71.4757) (2.39164, 71.0695) (2.39866, 70.6657)
            (2.40569, 70.2642) (2.41271, 69.8651) (2.41973, 69.4683) (2.42676, 69.0739)
            (2.43378, 68.6819) (2.4408, 68.2921) (2.44783, 67.9048) (2.45485, 67.5197)
            (2.46187, 67.137) (2.4689, 66.7566) (2.47592, 66.3786) (2.48294, 66.0028)
            (2.48997, 65.6294) (2.49699, 65.2582) (2.50401, 64.8894) (2.51104, 64.5228)
            (2.51806, 64.1585) (2.52508, 63.7966) (2.53211, 63.4368) (2.53913, 63.0794)
            (2.54615, 62.7242) (2.55318, 62.3713) (2.5602, 62.0206) (2.56722, 61.6722)
            (2.57425, 61.326) (2.58127, 60.9821) (2.58829, 60.6404) (2.59532, 60.3009)
            (2.60234, 59.9636) (2.60936, 59.6286) (2.61639, 59.2958) (2.62341, 58.9652)
            (2.63043, 58.6367) (2.63746, 58.3105) (2.64448, 57.9865) (2.65151, 57.6646)
            (2.65853, 57.345) (2.66555, 57.0275) (2.67258, 56.7121) (2.6796, 56.399)
            (2.68662, 56.088) (2.69365, 55.7791) (2.70067, 55.4724) (2.70769, 55.1679)
            (2.71472, 54.8654) (2.72174, 54.5651) (2.72876, 54.267) (2.73579, 53.9709)
            (2.74281, 53.677) (2.74983, 53.3852) (2.75686, 53.0955) (2.76388, 52.8079)
            (2.7709, 52.5224) (2.77793, 52.2389) (2.78495, 51.9576) (2.79197, 51.6783)
            (2.799, 51.4011) (2.80602, 51.126) (2.81304, 50.853) (2.82007, 50.582)
            (2.82709, 50.313) (2.83411, 50.0461) (2.84114, 49.7813) (2.84816, 49.5184)
            (2.85518, 49.2577) (2.86221, 48.9989) (2.86923, 48.7421) (2.87625, 48.4874)
            (2.88328, 48.2347) (2.8903, 47.984) (2.89732, 47.7353) (2.90435, 47.4886)
            (2.91137, 47.2438) (2.91839, 47.0011) (2.92542, 46.7603) (2.93244, 46.5215)
            (2.93946, 46.2847) (2.94649, 46.0498) (2.95351, 45.8169) (2.96054, 45.586)
            (2.96756, 45.3569) (2.97458, 45.1299) (2.98161, 44.9047) (2.98863, 44.6815)
            (2.99565, 44.4603) (3.00268, 44.2409) (3.0097, 44.0235) (3.01672, 43.8079)
            (3.02375, 43.5943) (3.03077, 43.3826) (3.03779, 43.1727) (3.04482, 42.9648)
            (3.05184, 42.7587) (3.05886, 42.5545) (3.06589, 42.3522) (3.07291, 42.1518)
            (3.07993, 41.9532) (3.08696, 41.7564) (3.09398, 41.5615) (3.101, 41.3685)
            (3.10803, 41.1773) (3.11505, 40.988) (3.12207, 40.8004) (3.1291, 40.6147)
            (3.13612, 40.4308) (3.14314, 40.2488) (3.15017, 40.0685) (3.15719, 39.89)
            (3.16421, 39.7134) (3.17124, 39.5385) (3.17826, 39.3654) (3.18528, 39.1941)
            (3.19231, 39.0246) (3.19933, 38.8568) (3.20635, 38.6908) (3.21338, 38.5266)
            (3.2204, 38.3641) (3.22742, 38.2034) (3.23445, 38.0444) (3.24147, 37.8872)
            (3.24849, 37.7317) (3.25552, 37.5779) (3.26254, 37.4259) (3.26957, 37.2755)
            (3.27659, 37.1269) (3.28361, 36.98) (3.29064, 36.8348) (3.29766, 36.6913)
            (3.30468, 36.5495) (3.31171, 36.4093) (3.31873, 36.2709) (3.32575, 36.1341)
            (3.33278, 35.999) (3.3398, 35.8656) (3.34682, 35.7338) (3.35385, 35.6037)
            (3.36087, 35.4752) (3.36789, 35.3483) (3.37492, 35.2231) (3.38194, 35.0996)
            (3.38896, 34.9777) (3.39599, 34.8573) (3.40301, 34.7387) (3.41003, 34.6216)
            (3.41706, 34.5061) (3.42408, 34.3922) (3.4311, 34.28) (3.43813, 34.1693)
            (3.44515, 34.0602) (3.45217, 33.9527) (3.4592, 33.8467) (3.46622, 33.7424)
            (3.47324, 33.6395) (3.48027, 33.5383) (3.48729, 33.4386) (3.49431, 33.3405)
            (3.50134, 33.2439) (3.50836, 33.1488) (3.51538, 33.0553) (3.52241, 32.9633)
            (3.52943, 32.8728) (3.53645, 32.7838) (3.54348, 32.6964) (3.5505, 32.6104)
            (3.55753, 32.526) (3.56455, 32.443) (3.57157, 32.3616) (3.5786, 32.2816)
            (3.58562, 32.2031) (3.59264, 32.1261) (3.59967, 32.0505) (3.60669, 31.9764)
            (3.61371, 31.9038) (3.62074, 31.8326) (3.62776, 31.7629) (3.63478, 31.6946)
            (3.64181, 31.6277) (3.64883, 31.5623) (3.65585, 31.4983) (3.66288, 31.4358)
            (3.6699, 31.3746) (3.67692, 31.3149) (3.68395, 31.2565) (3.69097, 31.1996)
            (3.69799, 31.1441) (3.70502, 31.0899) (3.71204, 31.0371) (3.71906, 30.9857)
            (3.72609, 30.9357) (3.73311, 30.8871) (3.74013, 30.8398) (3.74716, 30.7939)
            (3.75418, 30.7493) (3.7612, 30.7061) (3.76823, 30.6642) (3.77525, 30.6236)
            (3.78227, 30.5844) (3.7893, 30.5465) (3.79632, 30.51) (3.80334, 30.4747)
            (3.81037, 30.4408) (3.81739, 30.4081) (3.82441, 30.3768) (3.83144, 30.3467)
            (3.83846, 30.318) (3.84548, 30.2905) (3.85251, 30.2643) (3.85953, 30.2394)
            (3.86656, 30.2157) (3.87358, 30.1933) (3.8806, 30.1722) (3.88763, 30.1523)
            (3.89465, 30.1336) (3.90167, 30.1162) (3.9087, 30.1001) (3.91572, 30.0852)
            (3.92274, 30.0714) (3.92977, 30.059) (3.93679, 30.0477) (3.94381, 30.0376)
            (3.95084, 30.0288) (3.95786, 30.0211) (3.96488, 30.0146) (3.97191, 30.0093)
            (3.97893, 30.0053) (3.98595, 30.0023) (3.99298, 30.0006) (4, 30) (12.5, 30)
         };

		% recovery (complete)
			\addplot[name path = response, densely dashed] coordinates { % resilienceplot.py
				(0, 100) (1.4, 100) (1.4301, 99.985) (1.4602, 99.9401) (1.4903, 99.8653) (1.5204, 99.7603) (1.5505, 99.6253)
				(1.5806, 99.4602) (1.6107, 99.2649) (1.6408, 99.0394) (1.6709, 98.7836) (1.701, 98.4975) (1.7311, 98.1811)
				(1.7612, 97.8342) (1.7913, 97.4569) (1.8214, 97.0492) (1.85151, 96.6108) (1.88161, 96.1419) (1.91171, 95.6424)
				(1.94181, 95.1122) (1.97191, 94.5513) (2.00201, 93.9596) (2.03211, 93.3373) (2.06221, 92.6855) (2.09231, 92.0052)
				(2.12241, 91.2977) (2.15251, 90.564) (2.18261, 89.8053) (2.21271, 89.0228) (2.24281, 88.2176) (2.27291, 87.3909)
				(2.30301, 86.5438) (2.33311, 85.6774) (2.36321, 84.793) (2.39331, 83.8916) (2.42341, 82.9745) (2.45351, 82.0427)
				(2.48361, 81.0974) (2.51371, 80.1398) (2.54381, 79.171) (2.57391, 78.1921) (2.60401, 77.2044) (2.63411, 76.209)
				(2.66421, 75.2069) (2.69431, 74.1994) (2.72441, 73.1876) (2.75452, 72.1727) (2.78462, 71.1558) (2.81472, 70.1381)
				(2.84482, 69.1206) (2.87492, 68.1046) (2.90502, 67.0913) (2.93512, 66.0817) (2.96522, 65.077) (2.99532, 64.0783)
				(3.02542, 63.0869) (3.05552, 62.1038) (3.08562, 61.1302) (3.11572, 60.1673) (3.14582, 59.2162) (3.17592, 58.2781)
				(3.20602, 57.354) (3.23612, 56.4452) (3.26622, 55.5528) (3.29632, 54.678) (3.32642, 53.8219) (3.35652, 52.9856)
				(3.38662, 52.1703) (3.41672, 51.3772) (3.44682, 50.6074) (3.47692, 49.8621) (3.50702, 49.1423) (3.53712, 48.4493)
				(3.56722, 47.7842) (3.59732, 47.1482) (3.62742, 46.5423) (3.65753, 45.9678) (3.68763, 45.4258) (3.71773, 44.9175)
				(3.74783, 44.4439) (3.77793, 44.0064) (3.80803, 43.6059) (3.83813, 43.2436) (3.86823, 42.9208) (3.89833, 42.6385)
				(3.92843, 42.3979) (3.95853, 42.2001) (3.98863, 42.0464) (4.01873, 41.9377) (4.04883, 41.8738) (4.07893, 41.8536)
				(4.10903, 41.8757) (4.13913, 41.9392) (4.16923, 42.0426) (4.19933, 42.1849) (4.22943, 42.3648) (4.25953, 42.5811)
				(4.28963, 42.8326) (4.31973, 43.1182) (4.34983, 43.4365) (4.37993, 43.7865) (4.41003, 44.167) (4.44013, 44.5766)
				(4.47023, 45.0142) (4.50033, 45.4787) (4.53043, 45.9687) (4.56054, 46.4832) (4.59064, 47.0209) (4.62074, 47.5806)
				(4.65084, 48.1611) (4.68094, 48.7612) (4.71104, 49.3797) (4.74114, 50.0154) (4.77124, 50.667) (4.80134, 51.3335)
				(4.83144, 52.0136) (4.86154, 52.7061) (4.89164, 53.4098) (4.92174, 54.1234) (4.95184, 54.8459) (4.98194, 55.5759)
				(5.01204, 56.3124) (5.04214, 57.054) (5.07224, 57.7996) (5.10234, 58.5479) (5.13244, 59.2979) (5.16254, 60.0482)
				(5.19264, 60.7978) (5.22274, 61.5452) (5.25284, 62.2895) (5.28294, 63.0294) (5.31304, 63.7636) (5.34314, 64.4909)
				(5.37324, 65.2103) (5.40334, 65.9204) (5.43344, 66.6201) (5.46355, 67.3081) (5.49365, 67.9833) (5.52375, 68.6445)
				(5.55385, 69.2904) (5.58395, 69.9199) (5.61405, 70.5318) (5.64415, 71.1248) (5.67425, 71.6977) (5.70435, 72.2494)
				(5.73445, 72.7787) (5.76455, 73.2843) (5.79465, 73.765) (5.82475, 74.2197) (5.85485, 74.6472) (5.88495, 75.0462)
				(5.91505, 75.4155) (5.94515, 75.754) (5.97525, 76.0604) (6.00535, 76.3335) (6.03545, 76.5722) (6.06555, 76.7752)
				(6.09565, 76.9414) (6.12575, 77.0694) (6.15585, 77.1582) (6.18595, 77.2066) (6.21605, 77.2132) (6.24615, 77.177)
				(6.27625, 77.0967) (6.30635, 76.9712) (6.33645, 76.7991) (6.36656, 76.5794) (6.39666, 76.3108) (6.42676, 75.9921)
				(6.45686, 75.6221) (6.48696, 75.1997) (6.51706, 74.7238) (6.54716, 74.1985) (6.57726, 73.632) (6.60736, 73.0324)
				(6.63746, 72.408) (6.66756, 71.767) (6.69766, 71.1178) (6.72776, 70.4685) (6.75786, 69.8275) (6.78796, 69.2028)
				(6.81806, 68.6029) (6.84816, 68.0359) (6.87826, 67.5101) (6.90836, 67.0337) (6.93846, 66.615) (6.96856, 66.2623)
				(6.99866, 65.9837) (7.02876, 65.7876) (7.05886, 65.6821) (7.08896, 65.6756) (7.11906, 65.7762) (7.14916, 65.9923)
				(7.17926, 66.3291) (7.20936, 66.7794) (7.23946, 67.3326) (7.26957, 67.9779) (7.29967, 68.7046) (7.32977, 69.5021)
				(7.35987, 70.3597) (7.38997, 71.2667) (7.42007, 72.2124) (7.45017, 73.1862) (7.48027, 74.1772) (7.51037, 75.175)
				(7.54047, 76.1687) (7.57057, 77.1477) (7.60067, 78.1012) (7.63077, 79.0187) (7.66087, 79.8895) (7.69097, 80.7027)
				(7.72107, 81.4479) (7.75117, 82.1142) (7.78127, 82.691) (7.81137, 83.1678) (7.84147, 83.5416) (7.87157, 83.82)
				(7.90167, 84.0112) (7.93177, 84.1235) (7.96187, 84.1652) (7.99197, 84.1446) (8.02207, 84.0699) (8.05217, 83.9494)
				(8.08227, 83.7914) (8.11237, 83.6042) (8.14247, 83.3961) (8.17258, 83.1752) (8.20268, 82.95) (8.23278, 82.7287)
				(8.26288, 82.5195) (8.29298, 82.3307) (8.32308, 82.1706) (8.35318, 82.0475) (8.38328, 81.9697) (8.41338, 81.9454)
				(8.44348, 81.9829) (8.47358, 82.0894) (8.50368, 82.2632) (8.53378, 82.4987) (8.56388, 82.7902) (8.59398, 83.1317)
				(8.62408, 83.5178) (8.65418, 83.9425) (8.68428, 84.4002) (8.71438, 84.8851) (8.74448, 85.3915) (8.77458, 85.9136)
				(8.80468, 86.4458) (8.83478, 86.9822) (8.86488, 87.5172) (8.89498, 88.045) (8.92508, 88.5598) (8.95518, 89.056)
				(8.98528, 89.5278) (9.01538, 89.9694) (9.04548, 90.3751) (9.07559, 90.7392) (9.10569, 91.0559) (9.13579, 91.3225)
				(9.16589, 91.5424) (9.19599, 91.7202) (9.22609, 91.8605) (9.25619, 91.9678) (9.28629, 92.0467) (9.31639, 92.1017)
				(9.34649, 92.1373) (9.37659, 92.1582) (9.40669, 92.1688) (9.43679, 92.1736) (9.46689, 92.1773) (9.49699, 92.1844)
				(9.52709, 92.1994) (9.55719, 92.2268) (9.58729, 92.2712) (9.61739, 92.3372) (9.64749, 92.4293) (9.67759, 92.552)
				(9.70769, 92.7099) (9.73779, 92.9075) (9.76789, 93.149) (9.79799, 93.4334) (9.82809, 93.7555) (9.85819, 94.1099)
				(9.88829, 94.4915) (9.91839, 94.8949) (9.94849, 95.3148) (9.9786, 95.7459) (10.0087, 96.183) (10.0388, 96.6207)
				(10.0689, 97.0538) (10.099, 97.477) (10.1291, 97.885) (10.1592, 98.2725) (10.1893, 98.6342) (10.2194, 98.9648)
				(10.2495, 99.2591) (10.2796, 99.5117) (10.3097, 99.7174) (10.3398, 99.8709) (10.3699, 99.9668) (10.4, 100) (12.5, 100)
			};

		\path [spath/save=ln] (axis cs:4.16923, 42.0426) -- (axis cs:10.4, 100);
		\path [spath/transform={ln}{shift={(0,2em)}}, spath/split at keep start={ln}{.5}, spath/use=ln, spath/save=ar];
		\begin{pgfonlayer}{axis descriptions}
			\draw [spath/use=ar, -latex]
				node [pos = .5, above, sloped,align = center] (rpa) {resilient\\ policy adaption};
		\end{pgfonlayer}

		\fill[white] (rpa.north west) -- (rpa.south east) -- (rpa.north east) -- cycle;

         \path [spath/save = ar2] ($(axis cs:0, 30)!(spath cs:ar 0)!(axis cs:12.5, 30)$) -- ($(axis cs:0, 30)!(spath cs:ar 1)!(axis cs:12.5, 30)$);
         \draw [-latex, spath/arrow shortening=false, spath/transform={ar2}{shift={(0,-.5em)}}, spath/use=ar2]
            node [pos = .5, below, align = center] {pre-event\\ action policy};
		\end{axis}
	\end{tikzpicture}%
	\caption{Impact of a resilience event $D$ on the system performance and recovery through autonomous adaption of $F_t$ in a resilient system.}
	\label{fig:performance}
\end{figure}
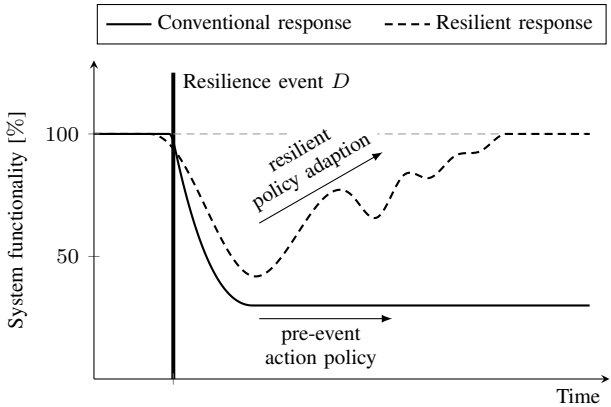

The primary aim of this paper is to analyze the implications \emph{unknown unknowns} have on the implementation of $F_t$.
We specifically focus on unanticipated, rare, and disruptive events, which will be
called \emph{resilience event} in the following.
This is the most relevant subset of unknown unknowns: An unanticipated event that is not rare is, essentially, anticipatable. Anticipated and rare events can be accommodated during system design, e.g., employing methods developed within the context of \cgls{urc}. A disruptive event has considerable impact on the system's operation, while a nondisruptive rare event is negligible.
We model resilience events as environmental state transitions with extremely low probability, i.e., $\Pr(E_{t+1} | E_{t}, A_t) < \varepsilon$ for some $\varepsilon > 0$. Since this $\varepsilon$ is very close to zero
and potentially non-computable with scientific methods, the asymptotic limit $\varepsilon\to 0$ is of particular interest.
We further assume that adapting to and recovering from a resilience event $D$ requires significantly different actions $A_t$ than during normal operation. That is, continuing to operate in the same way as prior to $D$ will lead to a considerable performance degradation. This is illustrated in \cref{fig:performance}.
Finally, we assume that the task performed by the system is sufficiently complex to require a nontrivial \cgls{ai}, i.e., $F_t$, $t\in\mathds N$ is implemented by a \cgls{lm}.

\subsection{Discussion} \label{sec:sysmod-d}
The connection of the system model to (deep) \cgls{rl} \cite{Sutton2020} and online convex optimization \cite{Shalev-Shwartz2012,Hazan2022} is rather obvious.
However, being consistent with Sutton's common model of the intelligent decision maker \cite{Sutton2022} and \cglspl{pomdp}, this model covers a much broader range of potential implementations.
For example, a conventional radio resource management system would neglect most of $H_t$ as well as the rewards $R_t$ to make scheduling decisions based on \cgls{csi}-reports and \cgls{qos}-demands, both part of the current observation $O_t$. As a second example, consider \cgls{dl} \cite{Goodfellow2016}, which covers a major share of modern \cgls{ml} techniques such as generative \cgls{ai}, \glspl{llm}, and transformer networks.\footnote{The authors are fully aware that this list contains redundancies.} The life-cycle of a \cgls{dl} system is roughly split into two phases, training and inference. During the training phase, a large amount of training examples is collected, often together with labels that represent the desired mapping the \cgls{lm} is supposed to learn. These examples are then used to train an algorithm, e.g., a deep \cgls{ann}. Once training is complete, the resulting algorithm is deployed and operates on previously unseen data, typically without any further training.
Mathematically, this is equivalent to embedding an omnipotent supervisor into the environment $E$. During a training phase, say for all $t\in[0:T_0]$, the supervisor isolates $S$ within $E$ and generates observation-reward tuples. Those are stored by $S$. After $t = T_0$, the \cgls{lm} uses this data set for training. From $t > T_0$, the system will be in deployment, using $O_t$ to infer $A_t$ and ignoring all further rewards.

\section{Statistical Machine Learning} \label{sec:sml}
The objective of system $S$ is to choose the mapping $F_t$ such that a function of the reward is maximized.
Consider implementing this through statistical \cgls{ml}. Then, $F_t$ corresponds to the model parameters $\theta$ that minimize a
risk functional
\begin{equation} \label{eq:risk}
	R(\theta) = \int_\Omega L(\omega, \theta) \Pr(\mathrm d\omega)
\end{equation}
over a model space $\Theta$ \cite{VonLuxburg2008,Vapnik1999}, where the loss function $L$ is $\Pr$-integrable and corresponds to a function of the negative reward. The integration in \cref{eq:risk} is with respect to a probability space $(\Omega, \mathscr A, \Pr)$. This space is (typically) not known explicitly and a primary objective of \cgls{ml} algorithms is to learn it from empirical data. 
Note that the majority of contemporary \cglspl{ai} are statistical \cgls{ml} systems.
In the following, we explore the limitations of this approach with respect to resilience events.

\subsection{Risk Minimization} \label{sec:riskmin}
Assume the probability space $(\Omega, \mathscr A, \Pr)$ is known and the only task is to minimize \cref{eq:risk}.
Consider a resilience event $D$ and let $\mathcal D$ be a subset of $\Omega$ relevant to the reaction and recovery from this event. Observe that, since
\begin{equation} \label{eq:markov}
	H_t \to A_t \to O_{t+1} \to R_{t+1},
\end{equation}
forms a Markov chain,
the transition probabilities $\Pr\left( E_{t+1} \,|\, E_{t}, A_{t} \right)$ are an integral component of $(\Omega, \mathscr A, \Pr)$. Recalling that a resilience event is characterized by extremely small $\Pr\left( E_{t+1} \,|\, E_{t}, A_{t} \right)$, it is reasonable to assume that $\Pr(\mathcal D) = \varepsilon$ for a small $\varepsilon > 0$.

The following proposition establishes that
\begin{equation}
	\int_\Omega L(\omega, \theta) \Pr(\mathrm d\omega) \approx \int_{\Omega\setminus \mathcal D} L(\omega, \theta) \Pr(\mathrm d\omega)
\end{equation}
for sufficiently small $\varepsilon$, i.e., minimizing \cref{eq:risk} \emph{does not} incentivise strategies to cope with resilience events.
Furthermore, since $\varepsilon$ is essentially non-computable \cite{Taleb2007}, regularization of $L$ to improve resilience appears to be, in general, infeasible.
\begin{proposition} \label{prop:risk}
	Let $(\Omega, \mathscr A, \Pr)$ be a probability space, $L$ be $\Pr$-integrable, and $\mathcal D\subset\Omega$ such that $\Pr(\mathcal D) = \varepsilon$. Then,
	\begin{equation}
		\int_{\mathcal D} L(\omega, \theta) \Pr(\mathrm d\omega) \to 0
	\end{equation}
	as $\varepsilon\to 0$ with linear convergence rate.
\end{proposition}
A proof of this proposition is presented in Appendix~\ref{sec:proofrisk}.
This result is hardly surprising to \cgls{ml} practitioners. It is also well known that $\Pr(D) = 0$ implies $\int_D L \,\mathrm d\mathds P = 0$. However, \cref{prop:risk} formally establishes well behaved convergence behavior around zero. This establishes without doubt that, regardless of the choice of $L$, a system design based on \cref{eq:risk} will not specifically implement functionalities to prepare for extremely rare events.

\subsection{Empirical Risk Minimization}
In many practical settings, the probability space $(\Omega, \mathscr A, \Pr)$ is unavailable and the risk functional \cref{eq:risk} is empirically approximated as
\begin{equation}
	R_\mathrm{emp}(\theta) = \frac{1}{\ell} \sum_{i=1}^\ell L(\omega_i, \theta),
\end{equation}
where $\{ \omega_1, \omega_2, \ldots, \omega_\ell \}$ are training samples drawn from $(\Omega, \mathscr A, \Pr)$. For sufficiently large $\ell$ and \cgls{iid} samples, $R_\mathrm{emp}$ converges to $R$ almost surely due to the strong \cgls{lln}. Thus, minimizing $R_\mathrm{emp}$ instead of $R$ is expected to result in a solution sufficiently close to the true optimum provided the data set is well chosen.

For example, consider the system $S$ employs \cgls{dl} to implement the \cgls{lm} as outlined in \cref{sec:sysmod-d}. \cGls{dl} operates under the assumption that the training data set is \cgls{iid}, i.e., $\mathcal T = \{ \omega_1, \omega_2, \ldots, \omega_\ell \}$ are realizations of the same distribution and pairwise independent. This implies permutation invariance of $\mathcal T$ with respect to the \cgls{dl} training process, i.e.,
any time-dependence information regarding state trajectories in $\mathcal T$ are discarded. The consequence is that the system implicitly assumes the Markov chain
\begin{equation} \label{eq:markovDL}
	H_{t-1} \to (O_t, R_t) \to A_t \to O_{t+1}
\end{equation}
instead of \cref{eq:markov}, i.e., it treats the environment in observation space as a first-order Markov chain, and implements a (measurable) mapping $O_t, R_t \mapsto A_t$. Note that, in general, \cref{eq:markovDL} is not implied by the environment being a Markov chain in $\mathcal E$ \cite{Kemeny1976}. Moreover, the \cgls{iid} assumption (and it being violated by most real-world data) is conjectured to be a primary reason for the brittleness of \cgls{dl} \cite{Kugelgen2024}.

Even statistical \cgls{ml} methods not relying on the \cgls{iid} assumption are subject to the challenges discussed in \cref{sec:riskmin}. However, in the \cgls{erm} case, it is possible to identify low probability events in the training set $\mathcal T$ and regularize the training objective to properly account for these tail probabilities \cite{Clemencon2025}. Clearly, this would require the relevant resilience events to be part of the training data set. This is unlikely to happen by chance due to their extreme low probability of occurrence. Recalling the \emph{unknown unknowns} discussion, synthetic data generation strategies to cover such events are, ultimately, a futile endeavour. This leaves online learning as the only viable statistical \cgls{ml} strategy to cope with resilience events.

\subsection{Online Learning} \label{sec:onlinelearning}
The generic system model in \cref{sec:sysmod} describes a system that learns continuously from its inputs $O_t, R_t$. Assume that training of the \cgls{lm} has converged for operation in the nominal environment. That is, the system consistently performs sufficiently optimal actions and is aware of this through a closed feedback loop. This nominal environment is a subset $\mathcal N$ of the environment $\mathcal E$ such that
\begin{equation} \label{eq:nominal}
	\Pr\{ E_{t+1}\in\mathcal N, E_t\in\mathcal N\} \approx 1.
\end{equation}
It is reasonable to assume that, in a properly engineered system, the observation space $\mathcal O$ is chosen such that there is negligible information loss in mapping $\mathcal N$ to $\mathcal O$.

\begin{figure}
	\centering
	\vspace*{5pt}
	\begin{tikzpicture}[
		  %every node/.style = {font = \footnotesize},
		  styleD/.style = {inner sep = 0, ellipse, minimum width = 6em, minimum height = 3em, densely dashed},
		  styleN/.style = {inner sep = 0, ellipse, minimum width = 4.5em, minimum height = 5.5em},
		  thick,
	   ]

	   % N and D
	   \node [draw, styleD] (setD) {};
	   \node [below=1.5em of setD, draw, styleN] (setN) {};

	   \node [anchor = south east] at (setN.south east) {$\mathcal{N}$};
	   \node [anchor = south east] at (setD.south east) {$\mathcal{D}$};

	   % Transitions
	   \path [name path = Dd1] (setD.east) -- (setD.west);
	   \path [name path = ND1] ($(setN.north west)!.3!(setN.south east)$) coordinate (Np1) |- (setD.north);
	   \path [name intersections={of=Dd1 and ND1}] (intersection-1) coordinate (Dp1);

	   \path [name path = ND2] ($(setN.north east)!.3!(setN.south west)$) coordinate (Np2) |- (setD.north);
	   \path [name intersections={of=Dd1 and ND2}] (intersection-1) coordinate (Dp2);

	   \draw[{Circle[open]}-{Stealth[] Circle}] (Np1) to[bend left]
		   node[left] (tp1) {$D$}
		   node[right] (tp2) {($\Pr(D) = \varepsilon$)}
		   (Dp1);
	   %\draw[{Circle[open]}-{Stealth[] Circle}] (Dp2) to[bend left] node[right] (tp2) {$\delta$} (Np2);

	   % Event space
	   \node [draw, rounded corners, inner xsep = 1em, fit = (setD) (setN) (tp1) (tp2)] (eventsp) {};
	   \node [rotate=90,above] at (eventsp.west) {Environment space $\mathcal{E}$};

	   % Projections
	   \node [right=8em of eventsp, draw, styleN] (projN) {};
	   \node [draw, styleD, rotate = 45] (projD) at (projN) {};

	   % Observation space
	   \node [draw, rounded corners, inner sep = .25em, fit = (projN)] (obssp) {};
	   \node [below, rotate = 90, align = center] at (obssp.east) {Observation space $\mathcal{O}$};

	   % observe
	   \draw [double, double distance = 3pt, -{Implies}, shorten <= .5em, shorten >= .5em] (eventsp) -- node [above=2pt] {Observation} node [below=2pt] {(projection)} (obssp);

	   % distribution
	   \draw [name path = Ndistr, shorten <= .5em, shorten >= 0em, -{Latex[]}] (setN.south east) .. controls (eventsp.south east) and ({$(eventsp.east)!.7!(obssp.west)$} |- eventsp.south) .. (obssp.south west);

	   \draw [name path = Ddistr, shorten <= .5em, shorten >= 0em, -{Latex[]}] (setD.20) .. controls (eventsp.north east) and ({$(eventsp.east)!.7!(obssp.west)$} |- eventsp.north) .. (obssp.north west);

	   % foo
	   \coordinate (tmp) at ($(eventsp.east)!.4!(obssp.west)$);
	   \path [name path = horiz] (current bounding box.north -| tmp) -- (current bounding box.south -| tmp);

	   \path [name intersections={of=horiz and Ndistr}] (intersection-1) node [anchor = north west]  {$\Pr$ on $\mathcal{N}$ induces $\mathds Q_{\mathcal{N}}$};
	   \path [name intersections={of=horiz and Ddistr}] (intersection-1) node [anchor = south west, inner ysep = 0]  {$\Pr$ on $\mathcal{N}$ induces $\mathds Q_{\mathcal{N}}$};
	\end{tikzpicture}%
	\caption{Illustration of the distributional shift from $\mathds{Q}_\mathcal{N} \to \mathds{Q}_\mathcal{N}$ in observation space $\mathcal O$ induced by a disruptive resilience event $D$.}
	\label{fig:distributionshift}
\end{figure}
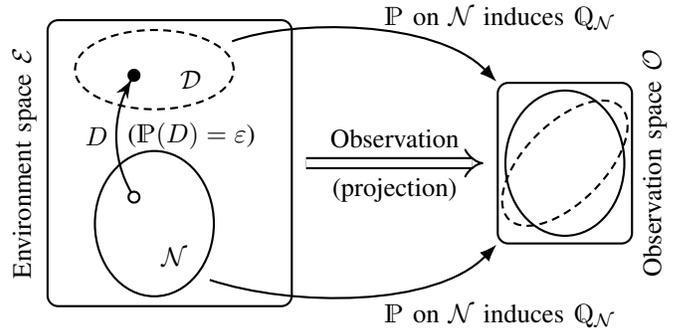

Consider a resilience event $D$ with major impact on the operating environment. In the state space $\mathcal E$, this is equivalent to transitioning out of the nominal environment $\mathcal N$ into a different region $\mathcal D \subset \mathcal E$ with $\mathcal N \cap \mathcal D = \emptyset$. Due to \cref{eq:nominal}, $\Pr\{ E_{t+1}\in\mathcal D, E_t\in\mathcal N\} = \varepsilon$, and for a disruptive and semi-permanent event, e.g., a natural disaster, $\Pr\{ E_{t+1}\in\mathcal N, E_t\in\mathcal D\} = \delta$ for small $\varepsilon, \delta > 0$. However, since the cardinality of $\mathcal O$ is significantly smaller than that of $\mathcal E$ and $\mathcal O$ approximately covers $\mathcal N$, $\mathcal N$ and $\mathcal D$ both map to overlapping regions of $\mathcal O$. Thus, the \cgls{lm} in $S$ perceives the resilience event $D$ as a distributional shift, say from $\mathds Q_\mathcal N$ to $\mathds Q_\mathcal D$.
See \cref{fig:distributionshift} for an illustruation.
Due to the No Free Lunch theorem \cite{Wolpert1996,Wolpert1997,Goodfellow2016,Simeone2023}, a stochastic \cgls{lm} optimal on $\mathds Q_\mathcal N$ is unlikely to also be optimal on $\mathds Q_\mathcal D$. As established before, learning these transition probabilities is integral to choosing reward-optimal actions $A_t$. Thus, some form of adaption to $\mathds Q_\mathcal D$ is necessary to continue operating well.

In the \cgls{rl} literature, the distributional shift from $\mathds Q_\mathcal N$ to $\mathds Q_\mathcal D$ is is treated under the umbrella of continual \cgls{rl} \cite{Khetarpal2022}, also known as lifelong \cgls{rl} \cite{Steinparz2022}, and non-stationary \cgls{rl} \cite{Moerland2023,Papoudakis2019}. This topic is at the forefront of contemporary \cgls{rl} research. A particularly noteworthy result form \cite{Steinparz2022} is that the widely employed \cgls{rl} methods \cgls{dqn} and \cgls{ppo} are unable to adapt to severe domain shifts such as those caused by resilience events.
The authors of \cite{Steinparz2022} propose a ``reactive exploration'' strategy that helps \cgls{ppo} cope with such shifts. It relies on detecting changes in the environmental distribution, followed by actively exploring the changed environment. However, ``exploration'' is nothing else than actively experimenting with the environment.
This, inevitably, leads to mistakes which may be expensive, irreversible, or even endanger human lives. A simple example would be the loss of communications in a critical moment during disaster relief, caused by the communication network exploring an unreliable mode of operation during its adaption process.
Moreover, it appears that allowing a complex \cgls{ai} system to freely explore its environment might be rather undesirable from the perspective of \cgls{ai} safety. 

Aside from safety aspects, current \cgls{rl} techniques for domain exploration, including \cite{Steinparz2022}, are not yet very data-efficient. Ultimately, this connects inherently to the fundamental, and still open, question of how much a machine can learn from passive observations alone \cite{LeCun2022}.

\section{Discussion} \label{sec:discussion}
We have established that continuous online learning is necessary for implementing resilient \cglspl{ran} based on statistical \cgls{ml}. Furthermore, we have observed that the current state-of-the-art in continual learning has only limited ability to deal with resilience events. A possible explanation for this, together with a solution approach, is considered in \cref{sec:CL}.
In \cref{sec:gpt}, we critically discuss our theoretical findings and potential limitations of our methodology with respect to experimental observations in \cglspl{llm}.

\subsection{Causal Learning} \label{sec:CL}
Consider the perspective provided in \cref{sec:onlinelearning} again. To keep the discussion simple,\footnote{This argument is easily extended to the general case.} consider an \cgls{ml} implementation that operates under the \cgls{iid} assumption. Then, the \cgls{lm} implements an observation-action mapping $O_t \mapsto A_t$, which either implicitly or explicitly depends on the state transition probabilities $p(O_{t+1}|O_t)$. Thus, it has learned some $p(A_t | O_t)$ on $\mathds Q_\mathcal N$ and needs to learn a new $q(A_t | O_t)$ on $\mathds Q_\mathcal D$. Thus, coping with resilience events is a \cgls{tl} problem.

Following the discussion in \cite[\S 13.2.3]{Simeone2023}, a positive information transfer from $p(O_{t+1}|O_t)$ to $q(A_t | O_t)$ is only possible if there exists a common set of features $u(O_t)$ invariant under the distributional shift, i.e., $p(O_{t+1}|u(O_t)) = q(O_{t+1}|u(O_t))$.
These features can be interpreted as causal effects, suggesting that moving from statistical to causal learning \cite{Scholkopf2022,Kugelgen2024} could significantly improve knowledge transfer and reduce the adaption time. Indeed, a successful transfer from $\mathds Q_\mathcal N$ to $\mathds Q_\mathcal D$ will require labeled data (e.g., rewards $R_{t+1}$ for explorative actions $A_t$) from the target domain, i.e., the system needs to observe and evaluate the effects of its actions on the changed environment. Thus, the system's adaption time is inherently connected to the system's explorative actions after $D$. This implies that improved knowledge transfer, e.g., through causal learning, directly corresponds to reducing the amount of required data collected under $\mathds Q_\mathcal D$. This establishes that the adaption time is not solely a matter of computational resources and motivates the recent advances towards causal \cgls{rl} \cite{Zeng2024}.

An important observation is that statistical online \cgls{ml} \emph{can} learn causal effects \cite{Schulte2024}, although structured discovery of causal effects is still its infancy \cite{Scholkopf2021}. Note, however, that there are strong indications that conventional training algorithms operating under the \cgls{iid} hypothesis, e.g., batch \cgls{sgd}, are unlikely to learn causal effects \cite{Kugelgen2024}. Another crucial benefit of using causal learning is the potential to infer the effects of actions without gathering experimental data from the changed environment \cite{Pearl2019}.

\subsection{GPTs and Reasoning Models} \label{sec:gpt}
Recent \cglspl{llm}, especially \gls{gpt}- and reasoning-models, exhibit tremendous capabilities. In light of those stunning results, one might be tempted to neglect the analysis in \cref{sec:sml} as misguided or, at least, overly pessimistic. We freely acknowledge that appropriately trained \cglspl{llm} might be able to deal with a large class of resilience events. Ultimately, however, the results in \cref{sec:sml} provide strong indication towards the opposite, i.e., that the most extreme resilience events require \cgls{ai} that relies on beyond-\cgls{ml} techniques. Indeed, it is straightforward to argue that resilience requires human-like problem solving capabilities.\footnote{Note that this is not the same as striving for an \cgls{agi}. Despite its complexity, operating a \cgls{ran} is still a narrowly defined task.} This is easily seen by observing that the origin of a resilience event might also be a deliberate malicious attack, supported by human ingenuity. It is this ingenuity, or commonsense, that is lacking in today's \cgls{ai}-models \cite{Larson2021,Shanahan2024}.
Still, modern \cglspl{llm} show remarkable \cgls{ood} generalization capabilities that are insufficiently explainable by statistical learning theory \cite{Reizinger2024a}, the foundation of the analysis in \cref{sec:riskmin}. This poses severe challenges to \cgls{ai} researchers and fuels the vigorous debate around the cognitive capabilities of contemporary \cglspl{ai} \cite{LeCun2022,Sutton2023,Guo2024c,Cui2024}. With respect to resilient \cglspl{ran}, further work towards understanding the extend to which \cgls{ood} generalization is sufficient to deal with the unknown unknowns is necessary.

\section{Conclusions} \label{sec:conclusions}
The ability to adapt to real-world complexity is a core capability of resilient systems. This includes challenges that are completely unanticipated during system design. We have emphasized the importance of these \emph{unknown unknowns} in resilience engineering and critically examined their implications on \cgls{ai} design for resilient \cglspl{ran}. Our results indicate strong limitations in widely employed \cgls{ml} techniques and suggest connections to continual learning, causal reasoning, and human-like commonsense. This positions resilient \cglspl{ran} at the forefront of \cgls{ai} research.

\appendices
\section{Proof of \cref{prop:risk}} \label{sec:proofrisk}
Let $(\Omega, \mathscr A, \Pr)$ be a probability space and $u$ a function integrable on this space. Consider an event $D\in\mathscr A$ with probability $\Pr(D) = \varepsilon$ for some small $\varepsilon > 0$. Let $\varepsilon_p$, $p\in\mathds N$, be a monotonically decreasing sequence $\varepsilon \ge \varepsilon_1 > \varepsilon_2 > \dots > 0$. Define the measures $\mu_p$, $p\in\mathds N$, as
\[ \mu_p(A) = \frac{\varepsilon_p}{\varepsilon} \Pr(A \cap D) + \frac{1-\varepsilon_p}{1-\varepsilon} \Pr(A \cap (\Omega\setminus D)) \]
for all $A\in\mathscr A$.
Observe that $(\mu_p(D))_{p\in\mathds N}$ is a monotonically decreasing sequence since
\begin{align*}
	\mu_p(D) &= \frac{\varepsilon_p}{\varepsilon} \Pr(D \cap D) + \frac{1-\varepsilon_p}{1-\varepsilon} \Pr(D \cap (\Omega\setminus D)) \\
	&= \frac{\varepsilon_p}{\varepsilon} \Pr(D) = \varepsilon_p.
\end{align*}
Trivially, $\mu_p$ is a measure on $(\Omega, \mathscr A)$. Furthermore, since
\begin{align*}
	\mu_p(\Omega) &= \frac{\varepsilon_p}{\varepsilon} \Pr(\Omega \cap D) + \frac{1-\varepsilon_p}{1-\varepsilon} \Pr(\Omega \cap (\Omega\setminus D)) \\
	&= \frac{\varepsilon_p}{\varepsilon} \Pr(D) + \frac{1-\varepsilon_p}{1-\varepsilon} \Pr(\Omega\setminus D) = 1,
\end{align*}
$(\Omega, \mathscr A, \mu_p)$ is a probability space for all $p$.

First, we establish that $u$ is $\mu_p$-integrable if it is $\Pr$-integrable.
Since $u$ is $\mathscr A$-measurable \cite[Def.~10.1]{Schilling2017}, there exist nonnegative $\mathscr A$-measurable functions $u^+, u^-$ such that $u = u^+ - u^-$ \cite[Cor.~8.12]{Schilling2017}.
By virtue of the sombrero lemma \cite[Thm.~8.8]{Schilling2017}, $u^+, u^-$ are each the pointwise limit of an increasing sequence of nonnegative simple functions $f^\pm_n$ measurable on $\mathscr A$, i.e., $u^\pm = \lim_{n\to\infty} f^\pm_n$. To simplify notation, consider an arbitrary $\Pr$-integrable nonnegative simple function $v$, with standard representations $v = \sum_{i=1}^N y_{i} \mathds 1_{A_{i}}$, where $N \in \mathds N$, $A_{i}$ are disjoint partitions of $\Omega$, and $y_{i}$ are nonnegative real numbers \cite[Def.~8.6]{Schilling2017}. Then,
\begin{align*}
	\MoveEqLeft[1]
	\int v \,\mathrm d\mu_p
	= \sum\nolimits_{i} y_i \int \mathds 1_{A_i}\,\mathrm d\mu_p
	= \sum\nolimits_{i} y_i \mu_p(A_i)
	\\&=
	\frac{\varepsilon_p}{\varepsilon} \sum\nolimits_{i}\! y_i \Pr(A_i \cap D) + \frac{1-\varepsilon_p}{1-\varepsilon}  \sum\nolimits_{i}\! y_i \Pr(A_i \cap (\Omega\setminus D))
	\\&=
	\frac{\varepsilon_p}{\varepsilon} \sum\nolimits_{i} y_i\!\int\! \mathds 1_{A_i \cap D} \mathrm d\mathds P + \frac{1-\varepsilon_p}{1-\varepsilon}  \sum\nolimits_{i} y_i \!\int\! \mathds 1_{A_i \cap (\Omega\setminus D)} \mathrm d\mathds P
	\\&=
	\frac{\varepsilon_p}{\varepsilon} \sum\nolimits_{i} y_i \int_{D} \mathds 1_{A_i} \mathrm d\mathds P + \frac{1-\varepsilon_p}{1-\varepsilon}  \sum\nolimits_{i} y_i \int_{\Omega\setminus D} \mathds 1_{A_i} \mathrm d\mathds P
	\\&=
	\frac{\varepsilon_p}{\varepsilon} \int_{D} v \,\mathrm d\mathds P + \frac{1-\varepsilon_p}{1-\varepsilon} \int_{\Omega\setminus D} v \,\mathrm d\mathds P.
\end{align*}
Hence, $f^\pm_n$ are $\mu_p$-integrable and, with the Monotone Convergence Theorem \cite[Thm.~9.6]{Schilling2017},
\begin{align*}
	\MoveEqLeft
	\int u^\pm \,\mathrm d\mu_p
	=  \lim_{n\to\infty} \int f_n^\pm\,\mathrm d\mu_p
	\\&=
	\frac{\varepsilon_p}{\varepsilon} \lim_{n\to\infty} \int_{D} f_n^\pm \,\mathrm d\mathds P  + \frac{1-\varepsilon_p}{1-\varepsilon} \lim_{n\to\infty} \int_{\Omega\setminus D} f_n^\pm \,\mathrm d\mathds P
\\&= \frac{\varepsilon_p}{\varepsilon} \int_{D} u^\pm \,\mathrm d\mathds P + \frac{1-\varepsilon_p}{1-\varepsilon}\int_{\Omega\setminus D} u^\pm \,\mathrm d\mathds P.
\end{align*}
Thus, $u^\pm$ are $\mu_p$-integrable and so is $u$.

For the limit $\varepsilon \to 0$, first observe that $\varepsilon_p \to 0$ as $p \to \infty$ faster than $\varepsilon$. Thus,
\begin{align*}
	\MoveEqLeft
	\lim_{p\to\infty} \int u \,\mathrm d \mu_p
	= \lim_{p\to\infty} \left\{ \frac{\varepsilon_p}{\varepsilon} \int_{D} u \,\mathrm d\mathds P + \frac{1-\varepsilon_p}{1-\varepsilon} \int_{\Omega\setminus D} u \,\mathrm d\mathds P \right\}
	\\&=
	\frac{\lim_{p\to\infty} \varepsilon_p}{\varepsilon} \int_{D} u \,\mathrm d\mathds P + \frac{1-\lim_{p\to\infty}\varepsilon_p}{1-\varepsilon} \int_{\Omega\setminus D} u \,\mathrm d\mathds P
	\\&=
	\frac{1}{1-\varepsilon} \int_{\Omega\setminus D} u \,\mathrm d\mathds P.
\end{align*}
Hence, $\int u \,\mathrm d\mu_p \to \int_{\Omega\setminus D} u \,\mathrm d\mathds P$ as $\varepsilon \to 0$.

For the convergence rate, define $I_X = \int_X u\, \mathrm d\mathds P$ and consider
\begin{multline*}
	r =
	\frac
	{\left| \frac{\varepsilon_{p+1}}{\varepsilon} I_D + \frac{1-\varepsilon_{p+1}}{1-\varepsilon} I_{\Omega\setminus D}- \frac{1}{1-\varepsilon} I_{\Omega\setminus D} \right|}
	{\left| \frac{\varepsilon_p}{\varepsilon} I_{D} + \frac{1-\varepsilon_p}{1-\varepsilon} I_{\Omega\setminus D} - \frac{1}{1-\varepsilon} I_{\Omega\setminus D} \right|}
	\\=
	\frac
	{\left| \frac{\varepsilon_{p+1}}{\varepsilon} I_{D} - \frac{\varepsilon_{p+1}}{1-\varepsilon} I_{\Omega\setminus D} \right|}
	{\left| \frac{\varepsilon_p}{\varepsilon} I_{D} - \frac{\varepsilon_p}{1-\varepsilon} I_{\Omega\setminus D} \right|}
	=
	\frac{\varepsilon_{p+1}}{\varepsilon_p}.
\end{multline*}
Since $\varepsilon_p > \varepsilon_{p+1}$, $r < 1$ for all $p$. Thus, the rate of convergence is linear \cite[p.~619]{Nocedal2006}. \hspace*{\fill}\IEEEQED

\balance
\bibliography{IEEEtrancfg,IEEEabrv,references}
\end{document}